\documentclass{mn2e}  
\usepackage{graphicx}

\def\sn{\hbox{S/N}}  
  
\def\vsin{\hbox{$v \sin i$}}  
  
\def\kms{\hbox{km\,s$^{-1}$}}

\def\em{\it}  
  
\def\degr{\hbox{$^\circ$}}

\def\omeq{\hbox{$\Omega_{\rm eq}$}}   
\def\dom{\hbox{$d\Omega$}}   
\def\kis{\hbox{$\chi^2$}}   
   
\def\drot{\hbox{differential rotation}}   
   
\def\xib{\hbox{$\xi$~Bootis~A}}

\def\hda{\hbox{HD~73350}}   
\def\hdb{\hbox{HD~76151}}   
\def\sco{\hbox{HD~146233}}   
\def\hdc{\hbox{HD~190771}}

\begin{document}  
  
\title[Toroidal vs. poloidal magnetic fields in Sun-like stars]  
{Toroidal vs. poloidal magnetic fields in Sun-like stars: a rotation threshold}
  
\makeatletter  
  
\def\newauthor{
  \end{author@tabular}\par  
  \begin{author@tabular}[t]{@{}l@{}}}  
\makeatother  
   
\author[P.~Petit et al.]  
{\vspace{1.5mm}
P.~Petit$^{1}$, B. Dintrans$^1$, S.K. Solanki$^2$, J.-F. Donati$^1$,  M. Auri\`ere$^1$, F. Ligni\`eres$^1$,  \\ 
{\hspace{-1mm}\vspace{1.5mm}\LARGE\rm J. Morin$^1$, F. Paletou$^1$, J.  Ramirez$^3$, C. Catala$^3$, R. Fares$^1$ }\\ 
$^1$Laboratoire d'Astrophysique de Toulouse-Tarbes, Universit\'e de Toulouse, CNRS, France\\ ({\tt petit@ast.obs-mip.fr, dintrans@ast.obs-mip.fr, donati@ast.obs-mip.fr, jmorin@ast.obs-mip.fr, fpaletou@ast.obs-mip.fr,}\\
{\tt auriere@ast.obs-mip.fr, ligniere@ast.obs-mip.fr, rim.fares@ast.obs-mip.fr})\\
$^2$Max-Planck Institut f\"ur Sonnensystemforschung, Max-Planck-Str. 2, 37191 Katlenburg-Lindau, Germany ({\tt solanki@mps.mpg.de})\\
$^3$LESIA, Observatoire de Paris-Meudon, 92195 Meudon, France\\
({\tt Claude.Catala@obspm.fr, jramirez@mesiog.obspm.fr})}

\date{$Revision: 1.18 $}  
\maketitle  
   
\begin{abstract}   
From a set of stellar spectropolarimetric observations, we report the detection of surface magnetic fields in a sample of four solar-type stars, namely \hda, \hdb, \sco\ (18 Sco) and \hdc.  Assuming that the observed variability of polarimetric signal is controlled by stellar rotation, we establish the rotation periods of our targets, with values ranging from 8.8~d (for \hdc) to 22.7 d (for \sco). Apart from rotation, fundamental parameters of the selected objects are very close to the Sun's, making this sample a practical basis to investigate the specific impact of rotation on magnetic properties of Sun-like stars.

We reconstruct the large-scale magnetic geometry of the targets as a low-order ($\ell<10$) spherical harmonics expansion of the surface magnetic field. From the set of magnetic maps, we draw two main conclusions. (a) The magnetic energy of the large-scale field increases with rotation rate. The increase of chromospheric emission with the mean magnetic field is flatter than observed in the Sun. Since the chromospheric flux is also sensitive to magnetic elements smaller than those contributing to the polarimetric signal, this observation suggests that a larger fraction of the surface magnetic energy is stored in large scales as rotation increases. (b) Whereas the magnetic field is mostly poloidal for low rotation rates, more rapid rotators host a large-scale toroidal component in their surface field. From our observations, we infer that a rotation period lower than $\approx$12 days is necessary for the toroidal magnetic energy to dominate over the poloidal component. \end{abstract}  
  
     
\section{Introduction}   
\label{sect:introduction}  

According to dynamo models, the variable magnetic field of the Sun is
the consequence of the interplay between two main ingredients. The first
ingredient is the vertical and latitudinal differential rotation that
succeeds at generating a large-scale toroidal magnetic field from an
initial poloidal field. The second ingredient is still a matter of
debate, with models invoking either the cyclonic convection in the
convection zone (Parker 1955) or the transport of decaying active
regions by meridional circulation (Dikpati et al. 2004) as possible
processes to regenerate the poloidal magnetic component. When acting
together, both effects succeed at building continuously a large-scale
magnetic field that oscillates with time, giving rise to the 22 yr
period of the solar cycle. Despite considerable progress in this field
since the very first solar dynamo model proposed by Parker (1955), there
are still many aspects of solar magnetism that the current models cannot
reproduce or did not thoroughly explore (see, e.g., the reviews of
Ossendrijver 2003 or Brandenburg \& Subramanian 2005). 

Our understanding of the solar dynamo can benefit from the observation of solar-type stars, where dynamo types marginal or inactive in the Sun can be observed, either because these analogues of the Sun are caught by chance in an unfrequented activity state or because their physical properties (in particular their mass and rotation rate) differ sufficiently from the Sun's to lead to a different dynamo output. Chromospheric emission, observed on stars possessing an extended convective envelope, has been monitored for decades as an indirect tracer of stellar magnetism (e.g. Baliunas et al., 1995) and offered a first opportunity to investigate the solar-stellar connexion. A variety of behaviours are observed, from erratic variations (common in young solar analogues) to smooth cycles (which seem typical of older stars like the Sun). Yet, the total chromospheric flux provides little information about the spatial organization of the magnetic field at the stellar surface, the key observable to collect for a comparison with magnetic geometries predicted by dynamo models.

Any useful effort at drawing a solar-stellar connexion from direct magnetic field measurements needs to address the case of solar twins, i.e. objects with stellar parameters very close to the Sun's, including a similar level of magnetic activity. Despite the strong scientific motivation, measuring a magnetic field on a strict twin of our Sun has always failed so far. Although being successful on a few objects quite different from the Sun by their mass (Valenti et al. 1996, Reiners \& Basri 2007) or rotation rate (R\"uedi et al. 1997, Donati et al. 2003, Petit et al. 2005), measurements of Zeeman signatures have always failed at detecting the magnetic field of low-activity solar twins, for well-identified reasons. 

A first option at our disposal, consisting in measuring the magnetic broadening of spectral lines (Saar 1988) is sensitive to the total photospheric magnetic flux but is plagued by the small fraction of the photosphere covered by strong magnetic fields. A second option consists in analyzing circularly polarized signatures inside line profiles (Donati et al. 1997) to recover information on the line-of-sight component of the magnetic vector. This second strategy suffers from the complex magnetic topology of cool stars, featuring a mixture of opposite polarities on the visible hemisphere, so that their respective polarized signatures cancel out whenever the rotational broadening is not sufficiently large to disentangle their respective contributions. For slow-rotators, only the largest-scale components of the surface field can add-up constructively to produce detectable circular polarization. On the Sun, this global component of the magnetic field displays a strength limited to a few Gauss only (Babcock \& Babcock 1955, Smith \& Balogh 1995, Sanderson et al. 2003, Mancuso \& Garzelli 2007). On other low-activity stars, the very tiny Zeeman signatures associated to such global fields have always escaped the scrutiny of observers so far.

We achieve this detection in the present study, with the help of spectropolarimetric data sets collected with the NARVAL spectropolarimeter. From a set of observations of a sample of four Sun-like stars covering a range of rotation periods, we reconstruct their large-scale photospheric magnetic geometry and discuss the impact of rotation on their magnetic properties.

\section{Stellar sample}
\label{sect:sample}

\begin{table*}
\caption[]{Fundamental parameters of the stellar sample. The effective temperature (T$_{\rm eff}$), surface gravity log(g), metallicity [M/H], mass, age and projected rotational velocity (\vsin) are taken from Valenti \& Fischer (2005), except the \vsin\ value of HD 146233 (see Sect. 3). The equatorial rotation period (P$_{\rm rot}^{\rm eq}$), difference in rotation rate between pole and equator (\dom) and inclination angle are derived from Zeeman-Doppler imaging.}
\begin{tabular}{cccccccccccccc}
\hline
Name & T$_{\rm eff}$ & log(g) & [M/H] & Mass & Age &  \vsin\ & {\bf P}${\bf _{\rm rot}^{\rm eq}}$ & \dom & inclination\\
 & (K) & [cm.s$^{-2}$] & [Sun] & (M$_\odot$) & (Gyr) & (\kms) & {\bf(d)} & (rad.d$^{-1}$) & (\degr)\\
\hline
Sun               & $5770$            & $4.44$                 & $0.00$                 & $1.0$                   & $4.3\pm1.7$             & $1.7$              & ${\bf25}$                  & 0.05 & --  \\
HD 146233 & $5791\pm 50$ & $4.41\pm 0.06$ & $0.03\pm 0.03$ & $0.98\pm 0.13$ & $4.7^{+2.7}_{-2.7}$ & $2.1\pm 0.5$ & ${\bf22.7\pm 0.5}$  & -- & $70^{+20}_{-25}$ \\
HD 76151    & $5790\pm 50$ & $4.55\pm 0.06$ & $0.07\pm 0.03$ & $1.24\pm 0.12$ & $3.6^{+1.8}_{-2.3}$ & $1.2\pm 0.5$ & ${\bf20.5\pm 0.3}$ & -- & $30\pm 15$ \\ 
HD 73350    & $5802\pm 50$ & $4.48\pm 0.06$ & $0.04\pm 0.03$ & $1.01\pm 0.14$ & $4.1^{+2.0}_{-2.7}$ & $4.0\pm 0.5$ & ${\bf12.3\pm 0.1}$ & $0.2\pm 0.2$ & $75^{+15}_{-20}$ \\ 
HD 190771  & $5834\pm 50$ & $4.44\pm 0.06$ & $0.14\pm 0.03$ & $0.96\pm 0.13$ & $2.7^{+1.9}_{-2.0}$ & $4.3\pm 0.5$ & ${\bf8.8\pm 0.1}$ & $0.12\pm0.03$ & $50\pm 10$ \\ 
\hline
\end{tabular}
\label{tab:param}
\end{table*}

Our stellar sample is constituted of four nearby dwarfs. Their physical
parameters are chosen to be as close as possible to the Sun's, except
for their rotation period (Table \ref{tab:param}). According to the
spectral classification study of Valenti \& Fischer (2005), all selected
objects have an effective temperature compatible with the solar value
within 1.3$\sigma$. They also possess a surface gravity compatible with
the Sun, except HD~76151 (lying 1.8$\sigma$ above the solar value). Two
of our targets are significantly over-metallic compared to the Sun, with
HD~190771 differing from the Sun by as much as 4.5$\sigma$. 

Evolutionary models matching these atmospheric parameters provide us with estimates of stellar ages and masses (also listed in Valenti \& Fischer, 2005). Stellar masses are all close to solar, except HD~76151 that lies 2$\sigma$ above 1 $M_\odot$. Note however that Nordstrom et al. (2004) propose for this star a mass of $0.92^{+0.09}_{-0.04}$ $M_\odot$, while all other stars of our list have consistent mass estimates in both catalogues. All ages are consistent with a solar age, but the related uncertainties are generally in excess of 2~Gyr. In particular, \hdc\ is probably younger than the Sun (given its metallicity and fast rotation), but is still sufficiently evolved to be taken as a probable main-sequence dwarf. As far as the dynamo itself is concerned, this large uncertainty in stellar ages could be a relevant issue in case the depth of the convection zone was expected to significantly vary during the course of the main sequence. $1 M_\odot$ models produced by stellar evolution codes predict however that the radius of the base of the convection zone remains constant within 5\% over the first 9 Gyrs of the main sequence (S. Th\'eado, private communication).

From this set of stellar parameters, and given the related uncertainties, we conclude that the internal structure of the selected objects are probably very close to the Sun's, with a dynamo action that should differ from the solar dynamo through the dominant effect of rotation, including differential rotation (and possibly an unknown meridional circulation). 

\section{Spectropolarimetric data set and magnetic mapping}
\label{sect:data}

\begin{table}
\caption[]{Journal of observations. From left to right, we list the
Julian date, the exposure time, the error bar in Stokes V LSD profiles
and the phase of the rotational cycle at which the observation was made
(calculated according to the rotation periods of Table 1 and imposing
JD=2454101.5 as phase zero for all stars).}
\begin{tabular}{ccccc}
\hline
Name & Julian Date & exp. time & $\sigma_{\rm LSD}$ & rot. phase \\
  & (2,450,000+) & sec. & $10^{-5}I_{c}$ & \\
\hline
HD 146233 & 4307.41 & 3600.0 & 5.894 & 0.0711 \\
 & 4309.36 & 3600.0 & 2.201 & 0.1571 \\
 & 4311.38 & 3600.0 & 2.029 & 0.2459 \\
 & 4312.37 & 3600.0 & 1.951 & 0.2894 \\
 & 4313.37 & 3600.0 & 3.395 & 0.3336 \\
 & 4315.37 & 3600.0 & 2.294 & 0.4216 \\
 & 4316.37 & 3600.0 & 2.174 & 0.4658 \\
 & 4322.37 & 3600.0 & 2.850 & 0.7298 \\
 & 4323.37 & 3600.0 & 3.234 & 0.7740 \\
 & 4331.35 & 3600.0 & 2.465 & 1.1258 \\
\hline
HD 76151 & 4122.51 & 2000.0 & 7.982 & 0.0250 \\
 & 4126.55 & 3200.0 & 4.887 & 0.2223 \\
 & 4127.56 & 3200.0 & 4.178 & 0.2712 \\
 & 4128.57 & 3200.0 & 3.727 & 0.3206 \\
 & 4130.56 & 3200.0 & 2.956 & 0.4176 \\
 & 4132.55 & 3200.0 & 3.457 & 0.5149 \\
 & 4133.59 & 3200.0 & 3.570 & 0.5655 \\
 & 4134.56 & 3200.0 & 3.335 & 0.6129 \\
 & 4135.58 & 3200.0 & 4.114 & 0.6626 \\
 & 4136.55 & 3200.0 & 4.404 & 0.7100 \\
 & 4146.58 & 3200.0 & 3.954 & 1.1993 \\
 & 4150.56 & 3200.0 & 3.471 & 1.3933 \\
 & 4159.46 & 3200.0 & 3.573 & 1.8277 \\
\hline
HD 73350 & 4127.47 & 2400.0 & 7.113 & 0.1122 \\
  & 4128.43 & 2400.0 & 6.274 & 0.1902 \\
 & 4130.43 & 2400.0 & 5.329 & 0.3518 \\
 & 4133.45 & 2400.0 & 8.391 & 0.5983 \\
 & 4134.47 & 2400.0 & 6.899 & 0.6813 \\
 & 4135.45 & 2400.0 & 4.943 & 0.7605 \\
 & 4136.42 & 2400.0 & 5.519 & 0.8392 \\
 & 4138.42 & 3600.0 & 5.659 & 1.0021 \\
 & 4140.43 & 2400.0 & 5.569 & 1.1656 \\
\hline
HD 190771 & 4307.52 & 2400.0 & 4.584 & 0.4121 \\
 & 4308.47 & 2400.0 & 4.313 & 0.5194 \\
 & 4309.50 & 2400.0 & 3.506 & 0.6369 \\
 & 4310.52 & 2400.0 & 3.159 & 0.7528 \\
 & 4311.52 & 1600.0 & 4.230 & 0.8663 \\
 & 4312.50 & 1600.0 & 4.055 & 0.9783 \\
 & 4313.50 & 1600.0 & 5.451 & 1.0909 \\
 & 4315.51 & 1600.0 & 4.439 & 1.3199 \\
 & 4316.51 & 1600.0 & 4.221 & 1.4332 \\
 & 4321.48 & 1600.0 & 4.772 & 1.9983 \\
 & 4322.50 & 1600.0 & 5.270 & 2.1146 \\
 & 4323.50 & 1600.0 & 5.792 & 2.2281 \\
 & 4327.50 & 1600.0 & 4.615 & 2.6830 \\
 & 4331.43 & 1600.0 & 4.116 & 3.1291 \\
\hline
\end{tabular}
\label{tab:journal}
\end{table}

\begin{figure*}
\centering
\mbox{\includegraphics[width=4cm]{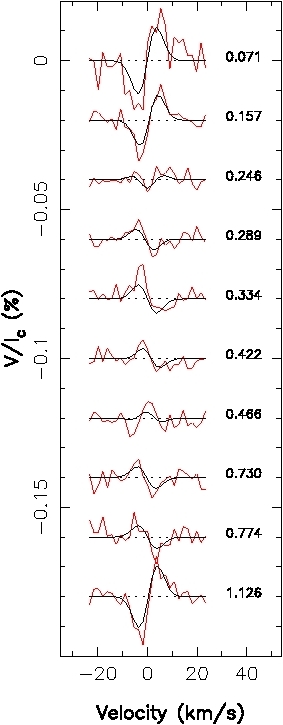}
\includegraphics[width=4cm]{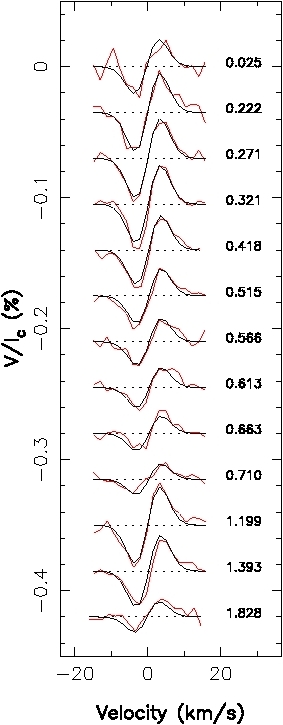}
\includegraphics[width=4cm]{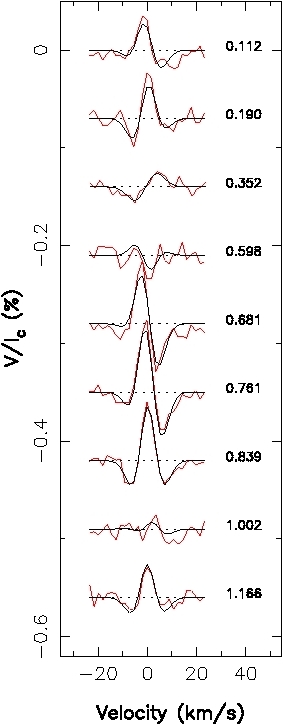}
\includegraphics[width=4cm]{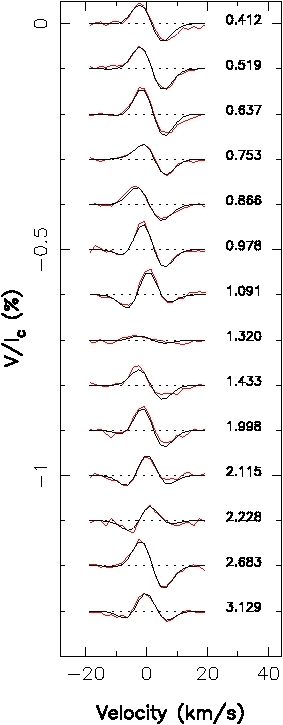}}
\caption{From left to right, time-series of Stokes V profiles of HD 146233, HD 76151, HD 73350 and HD 190771, in the rest-frame of the star. Red lines
represent the data and black lines correspond to synthetic profiles of our magnetic model. Successive profiles are shifted vertically for display clarity and rotational phases of observations are indicated in the right part of the plot.}
\label{fig:stokesV}
\end{figure*}

The observational material consists of high-resolution spectra obtained simultaneously in classical spectroscopy (Stokes I) and circularly polarized light (Stokes V) in 2007 winter and summer, using the newly installed NARVAL stellar spectropolarimeter at T\'elescope Bernard Lyot (Observatoire du Pic du Midi, France). As a strict copy of ESPaDOnS (Petit et al. 2003), NARVAL provides full coverage of the optical domain (370 nm to 1,000 nm) in a single exposure, at a resolving power of 65,000, with a peak efficiency of about 15\% (telescope and detector included). It is constituted of a bench-mounted spectrograph (based on a dual-pupil design and stored in a double thermal enclosure), fiber-fed from a Cassegrain-mounted module containing all polarimetric facilities. A series of 3 Fresnel rhombs (two half-wave rhombs that can rotate about the optical axis and one quarter-wave rhomb) are employed to perform a very efficient polarimetric analysis over the whole spectral domain. They are followed by a Wollaston prism which splits the incident light into two beams, respectively containing light linearly polarized perpendicular/parallel to the axis of the prism. The two beams produced by the Wollaston prism are imaged onto the two optical fibers that carry the light to the spectrograph. Each Stokes V spectrum is obtained from a combination of four sub-exposures taken with the half-wave rhombs oriented at different azimuths (Semel et al. 1993). The data reduction is performed by Libre-Esprit, a dedicated, fully automated software described by Donati et al. (1997) and implementing the optimal spectral extraction principle of Horne (1986) and Marsh (1989).

\begin{figure}
\centering
\includegraphics[width=9cm]{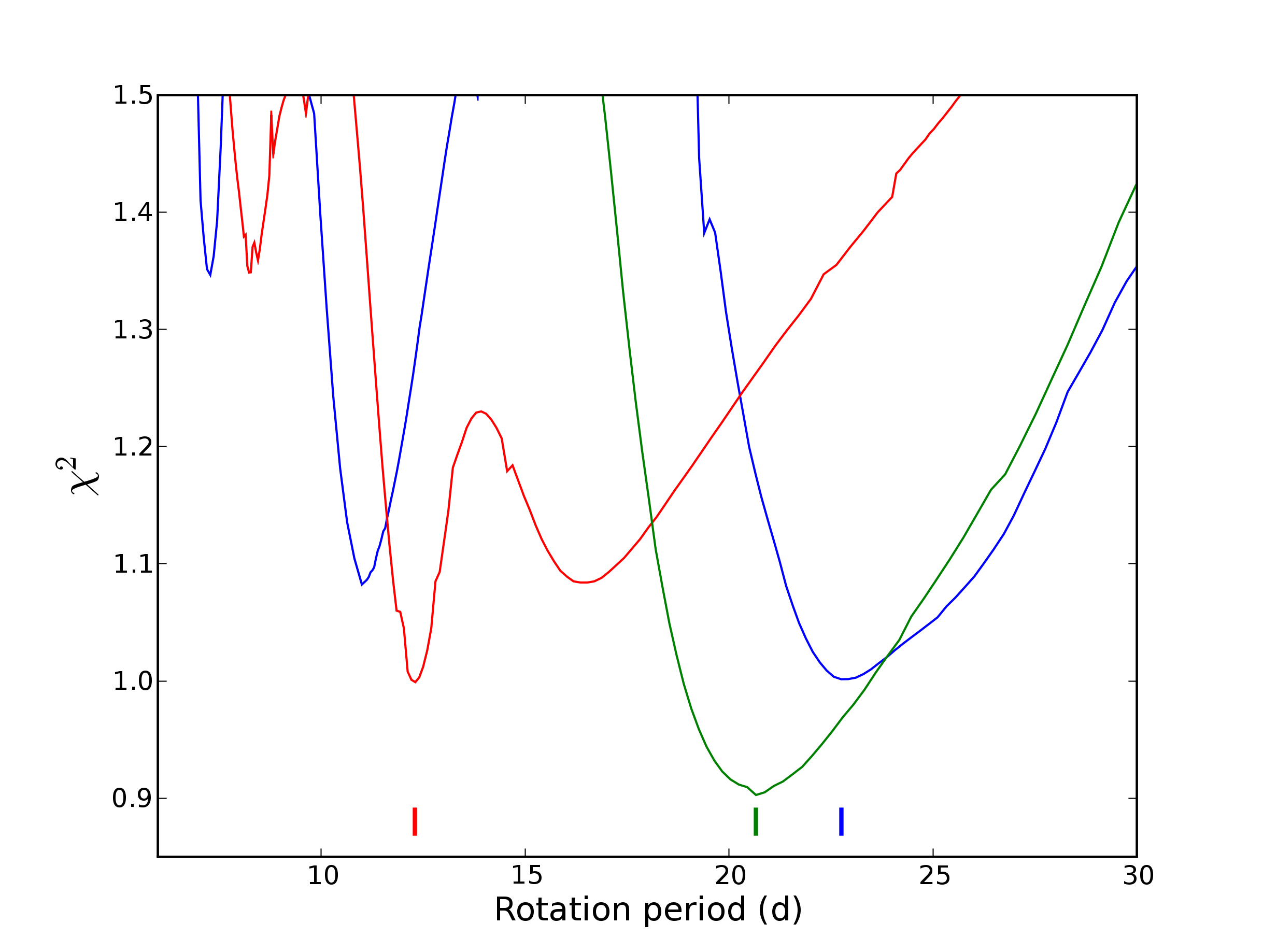}
\caption{Periodograms obtained for HD 73350, HD 76151 and HD 146233 (red, green and blue lines, respectively).}
\label{fig:period}
\end{figure}

\begin{figure}
\centering
\includegraphics[width=6cm]{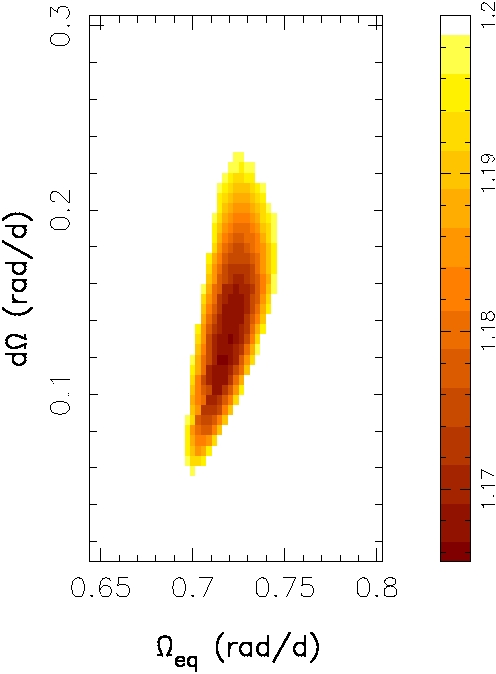}
\caption{\kis\ map obtained when varying the differential rotation parameters of HD 190771. \omeq\ is the rotation rate of the equator and \dom\ represents the difference in rotation rate between pole and equator (assuming a $\sin ^2$ dependence of rotation rate with colatitude).}
\label{fig:drot}
\end{figure}

A total of 9 to 14 spectra was collected for each star, at a rate of one
spectrum every clear night (Table \ref{tab:journal}). A single, average
photospheric line profile was extracted from each spectrum using the LSD
technique (Donati et al. 1997), according to a line-list matching a
solar photospheric model. Using a total of some 5,000 atomic spectral
lines  with wavelengths between 370~nm and 1~$\mu$m, the noise level of
the mean Stokes V profiles is reduced by a factor of about 40 with
respect to the initial spectrum. The resulting noise level, listed in
Table \ref{tab:journal}, ranges between $2\times 10^{-5}$ and  $8\times
10^{-5}$ $I_{\rm c}$ (where $I_{\rm c}$ denotes the continuum level)
depending on the source, on the fluctuating weather and on the adopted
exposure time. The recorded sets of Stokes V profiles are plotted in
Fig.~\ref{fig:stokesV}. Circularly polarized signals are detected for
each star, centered at the radial velocity of the unpolarized mean line
profile. We interpret this spectral line polarization as Zeeman
signatures related to the presence of large-scale photospheric magnetic
fields. This detection, achieved for the first time in solar twins, is
made possible by the combination of the large spectral domain (ensuring
highly efficient multi-line techniques) and high \sn\ of polarized
spectra (achievable thanks to the high collecting power of the
instrumental device).

Assuming that the observed temporal variability of Stokes V profiles is controlled by the stellar rotation, we reconstruct the magnetic geometry of our targets by means of Zeeman-Doppler imaging (ZDI). We employ here the modelling strategy of Donati \& Brown (1997), including also the spherical harmonics expansion of the surface magnetic field implemented by Donati et al. (2006) in order to easily distinguish between the poloidal and toroidal components of the reconstructed magnetic field distribution. We use a linear limb-darkening coefficient equal to 0.75 (Neckel 2003, but note that the imaging procedure is mostly unsensitive to small variations of this parameter) and we adopt the \vsin\ values derived by Valenti \& Fischer (2005), except for HD~146233 for which we choose \vsin=2.1 \kms\ (since the estimate of Valenti \& Fischer is not compatible with the rotation period we derive for this star). Prior to rotational broadening, we adopt a very simple gaussian model for the unpolarized line profile, in which any type of line broadening other than rotational enter, including instrumental smearing. We only have to slightly adjust the FWHM of the local profile (between 10.1 and 10.4 \kms, depending on the star) to obtain a convincing adjustment of the observed Stokes I LSD profiles.

To determine the rotation period of our targets, we use the principle of
maximum entropy image reconstruction and calculate a set of magnetic
maps for each star, assuming various values of the rotation period
(Fig.~\ref{fig:period}). We then adopt the rotation period that
minimizes the \kis\ of the reconstructed spectra, at fixed information
content (following the approach of Petit et al. 2002). For all stars
except \hdc, the temporal evolution of the polarized signal is
consistent with a solid-body rotation. For \hdc\ however, this
assumption does not yield a convincing adjustment of the Stokes V
time-series (\kis\ = 1.5). A better fit to the profiles (\kis\ = 1.1) is
obtained by including a solar-like \drot\ law in the imaging procedure
(Petit et al., 2002 and Fig.~\ref{fig:drot}). Using the same method with
HD 73350, we also obtain a \kis\ minimum in the \omeq-\dom\ plane, but
with error bars too large to exclude a solid-body rotation. The two
other stars do not show any \kis\ minimum along the \dom\ axis. 

For HD 73350, HD 76151 and HD 190771, our estimates of the rotation
period (see Table \ref{tab:param}) are 20\% to 25\% larger than
literature values derived from the less accurate period determination
using chromospheric activity indicators (Noyes et al. 1984, Wright et
al. 2004). For HD 146233, the level of chromospheric emission and the
periodic modulation of Zeeman signatures provide consistent estimates.
The observed rotation phases are listed in Table \ref{tab:journal},
using the rotation periods of Table \ref{tab:param} and imposing
JD=2454101.5 (2007 Jan. 01 at 00:00 UT) as phase zero for all stars.

From the combined knowledge of the rotation period and \vsin, we can
finally deduce the inclination angle $i$ of the rotation axis with
respect to the line of sight (Table \ref{tab:param}), assuming that
for all targets the stellar radius can be taken equal to the solar
radius (luminosities and effective temperatures taken from Valenti \&
Fischer, 2005, ensure that stellar radii of our targets are equal to the
Sun's within $\approx$5\%).  Using this set of stellar parameters, we
then reconstruct photospheric magnetic maps of our stellar sample
(Figs.~ \ref{fig:map1} and \ref{fig:map2}). By doing so, all data sets
are fitted down to the noise level ($0.9\leq \chi^2 \leq 1.1$, depending
on the star). We note that the hypothesis of a purely potential field is
actually sufficient to adjust the Stokes V profiles of HD 146233 and HD
76151 at the same \kis, without affecting the magnetic map. For HD 73350
however, assuming a purely potential magnetic field does not allow the
inversion procedure to reach better than \kis=1.5. The situation is even
worse with HD 190771, for which the final \kis\ cannot be better than 5
without the addition of a toroidal magnetic field. We check that the
magnetic maps remain essentially unaffected by the $\ell$ limit,
provided that the magnetic expansion respects $\ell \ge 4$ for HD 73350,
$\ell \ge 3$ for HD 190771, $\ell \ge 2$ for HD 146233 and HD 76151. For
all four stars, we choose to limit the magnetic model to a spherical
harmonics expansion of degree $\ell \le 10$.  

From the estimated spherical harmonics coefficients, we can derive a set
of numerical quantities to characterize the reconstructed magnetic
topologies (Table \ref{tab:dynamo}). To evaluate the uncertainties on
these numerical estimates, we vary the values of stellar parameters
(rotation period, \vsin, inclination angle) over the width of the error
bar on each individual parameter and reconstruct a magnetic map for each
new combination of the parameters. The observed variations in the output
magnetic quantities are quoted as error bars in Table \ref{tab:dynamo}
and give an estimate of the uncertainties on the reconstructed magnetic
geometries. 

\begin{figure*}
\centering
\mbox{\includegraphics[width=8cm]{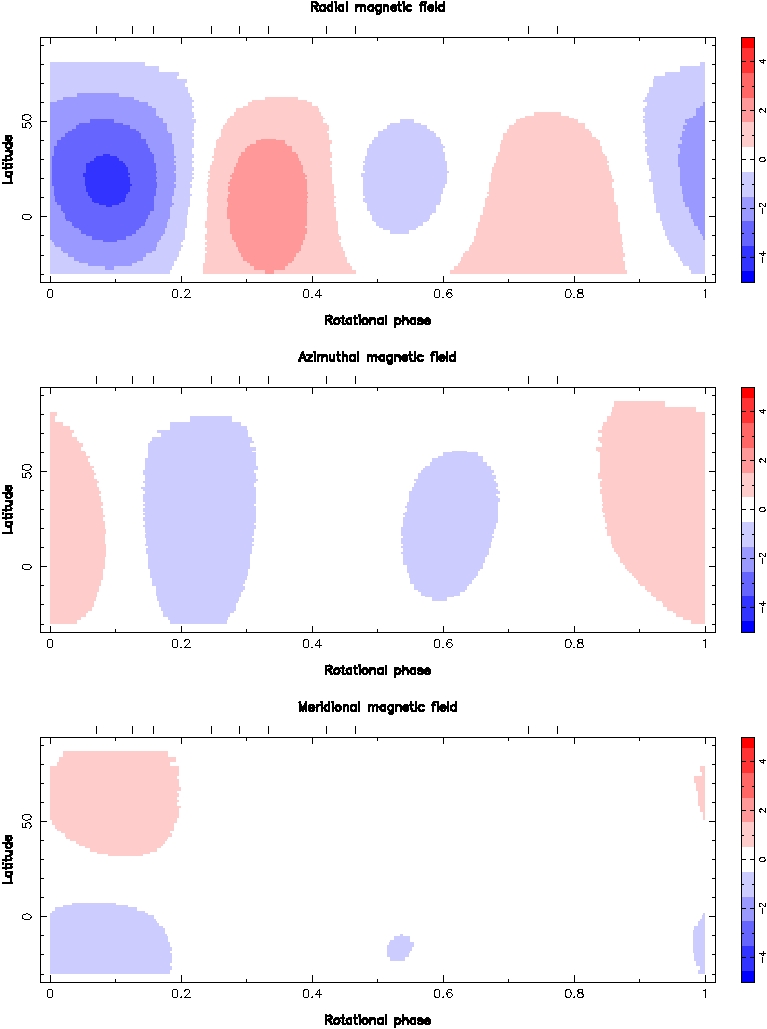}
\vspace{1cm}
\includegraphics[width=8cm]{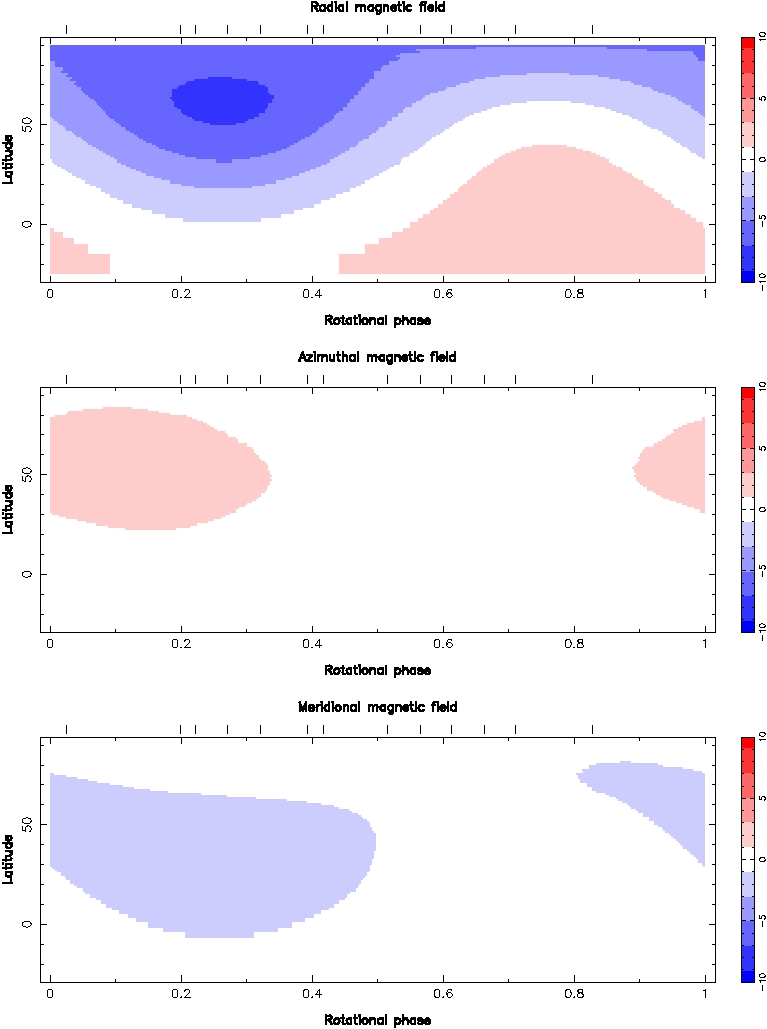}}
\caption{Magnetic maps of HD 146233 and HD 76151 (left and right panel, respectively). Each chart illustrates the field projection onto one axis of the spherical coordinate frame. The magnetic field strength is expressed in Gauss. Vertical ticks above the charts indicate the observed rotational phases. Note that color scales are not the same for every star.}
\label{fig:map1}
\end{figure*}

\begin{figure*}
\centering
\mbox{\includegraphics[width=8cm]{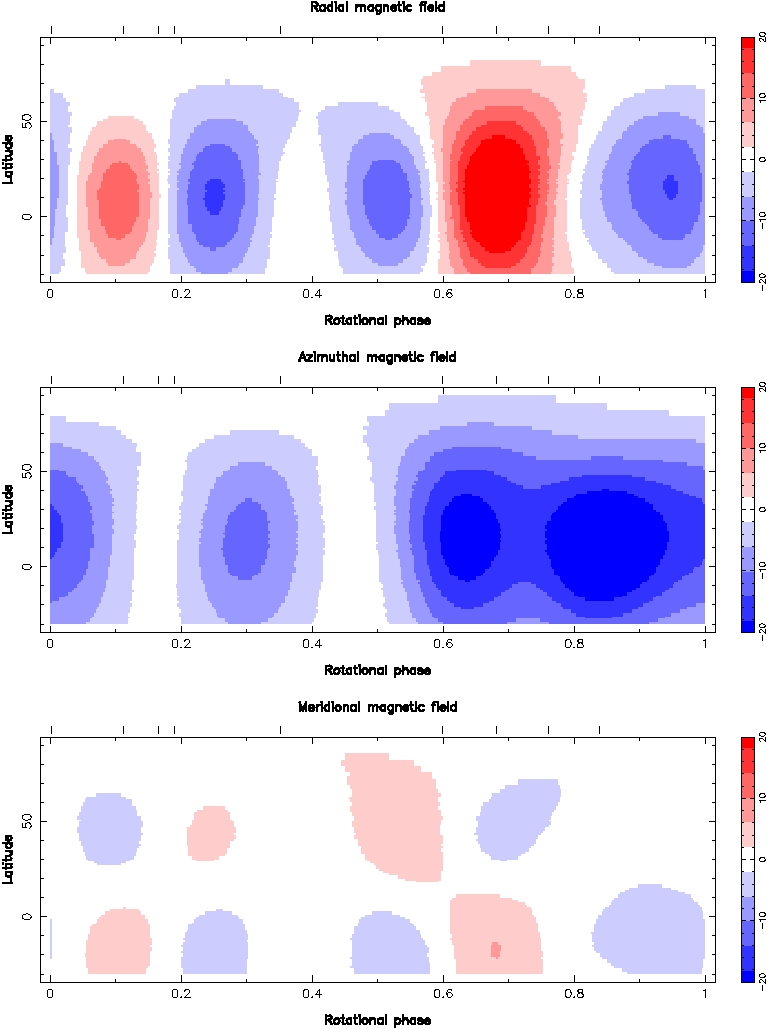}
\vspace{1cm}
\includegraphics[width=8cm]{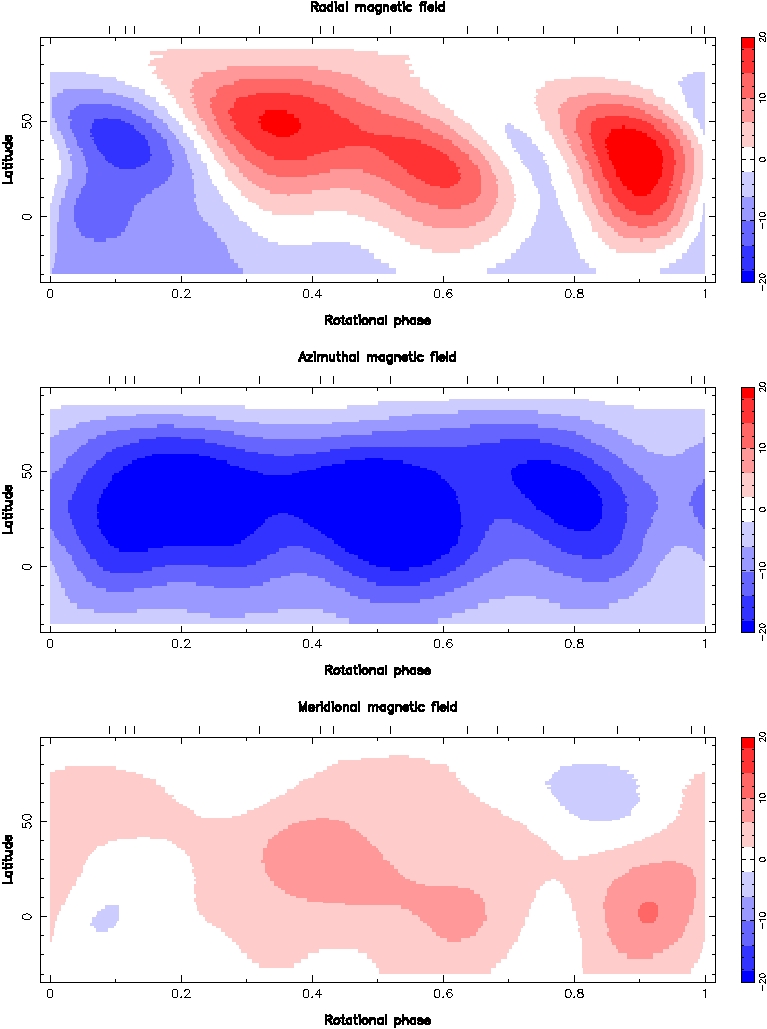}}
\caption{Same as Fig.~\ref{fig:map1}, for HD 73350 (left panel) and HD
190771 (right panel).}
\label{fig:map2}
\end{figure*}

\section{Discussion}

\subsection{Magnetic energy of the large-scale field}
\label{sect:energy}

The mean (unsigned) magnetic field of the reconstructed maps increases
with the rotation rate (Figs.~\ref{fig:map1}, \ref{fig:map2} and
\ref{fig:rhk}  and Table \ref{tab:dynamo}). This trend is in qualitative
agreement with previous studies investigating the rotational dependence
of various magnetic tracers, in particular the chromospheric flux (e.g.
Noyes et al. 1984, Wright et al. 2004, Hall et al. 2007a) or the Zeeman
broadening of magnetically-sensitive spectral lines (Saar 2001).
From our set of spectropolarimetric observations, we derive that the
magnetic energy of the large-scale field increases by a factor of
$\approx120$ from HD~146233 (P$_{\rm rot}^{\rm eq}$ = 22.7~d) to
HD~190771 (P$_{\rm rot}^{\rm eq}$ = 8.8~d). This evolution affects both
poloidal and toroidal magnetic components, but the poloidal magnetic
energy grows by a factor of about 40, while the toroidal magnetic energy
is increased by a factor of $10^4$.

From our sets of spectra, we can estimate the chromospheric emission of
the selected stars from the $R'_{\rm HK}$ parameter (defined as the
ratio of the chromospheric emission in CaII H \& K line cores to the
stellar bolometric emission). Plotted against the mean magnetic field
(Fig.~\ref{fig:rhk}), the chromospheric emission is following a
power-law of the form $\log (R'_{\rm HK}) = 0.32 \log(B_{\rm
mean}) - 4.98$. We obtain a very similar result when using
literature values of $R'_{\rm HK}$ instead of NARVAL values ($R'_{\rm HK} \propto B_{\rm mean}^{0.33}$). This dependence differs
from solar observations, from which an increase like $B_{\rm
mean}^{0.6}$ is derived in the quiet Sun and active regions (Schrijver et al.
1989, Ortiz \& Rast 2005).

In the case of low \vsin\ values, Zeeman-Doppler magnetic maps are only sensitive to low-order field components, while the measured chromospheric flux includes the contribution of smaller-scale elements as well (active regions and plages), irrespective of the rotation rate. Since the spatial resolution achieved through ZDI increases with \vsin, one could then argue that higher-order magnetic components might be resolved in the most rapidly-rotating stars of our sample and might thus be included in the magnetic energy we measure, explaining why the mean magnetic field seems to grow much faster with rotation than the chromospheric flux. We actually observe that spherical harmonics modes with $\ell \le 3$ (spatially resolved for all stars of our sample) always contain more than 94\% of the total photospheric magnetic energy, even for stars with the largest \vsin, and despite the fact that our magnetic modelling allows for higher-order magnetic components to be reconstructed up to $\ell = 10$. In fact, the \vsin\ range of our sample (from 1.2 to 4.3 \kms) is sufficiently limited to ensure a consistent behaviour of the imaging procedure for all targets (given the spectral resolution of NARVAL, limited to 65,000 in polarimetric mode), and therefore a consistent estimate of the magnetic energy of the large-scale field.

From these arguments, we propose that the observed discrepancy with the
solar case is genuine. A first possibility is that horizontal magnetic
fields (in particular the prominant toroidal field components seen in
Fig.~\ref{fig:map2}) are not able to penetrate the chromosphere and
contribute to produce enhanced CaII emission. To test this idea, we have
calculated the power law we obtain by considering the radial field
component only (Fig.~\ref{fig:rhk}, blue symbols). By doing so, we only
slightly change the exponent (with $R'_{\rm HK} \propto B_{\rm mean}^{0.36}$) and remain still very far from the steeper solar
dependence. A second possibility is that a higher coverage of cool spots
in fast-rotating Suns may contribute to alter the exponent in the
power-law (which is valid for the quiet Sun and plages, not for
sunspots). However, the dominant contributor to the Zeeman signatures we
observe is probably not originating from cool spots (except possibly the spot penumbra), since the contrast with
the rest of the photosphere will mostly hide the signal formed in these
dark regions. By constraining the temperature of magnetic regions
observed in cool active stars, previous studies (R\"uedi et al. 1997,
Petit et al. 2005) confirm that the observed magnetic elements are
hotter than the surrounding (non magnetic) material. Petit et al. (2005)
also show that Stokes V Zeeman signatures formed by the large-scale
field of solar-type stars are asymmetric, with a difference of area and
amplitude between the two lobes of the profile typical of solar faculae.

We therefore propose a third interpretation and suggest that
a larger fraction of the surface magnetic energy is stored in large
scales as rotation grows, so that a larger fraction of the actual
magnetic energy becomes measurable through ZDI, even when the field
reconstruction is limited to large-scale structures. This conclusion is also supported by the observed increase of the mean magnetic field, for which a power-law of the form  $B_{\rm mean} = 2.8 \times 10^{3} P_{\rm rot}^{-1.8}$ is derived from our observations. This increase is steeper than that derived from Zeeman broadening of spectral lines (a measurement technique also sensitive to smaller-scale magnetic elements, see Sect. \ref{sect:introduction}), from which Saar (2001) obtains an exponent equal to -1.2.

Finally, we note that the observed increase of
the mean magnetic field with the rotation rate is also in agreement
with recent 3-D MHD simulations of fast-rotating Suns (Brown et
al. 2007a, 2007b). In these simulations, the global field
becomes stronger within the bulk of the convection zone when the
rotation rate is increasing. This behaviour contrasts
with classical solar MHD models where weak magnetic fields with little
global-scale structure are produced and where fluctuating parts dominate near
the surface (Brun et al. 2004). In other words, a way to organize and
amplify magnetic fields in these global MHD simulations would consist in
increasing the star rotation, what we also observe here with these four
Sun-like stars.

\subsection{Large-scale toroidal component}

The second noticeable effect of rotation is to increase the fraction of
the magnetic energy stored into a large-scale toroidal component of the
surface magnetic field (Figs.~\ref{fig:map1} and \ref{fig:map2}). While
the toroidal component is negligible for HD~146233 and HD~76151
(representing respectively 1\% and 7\% of their large-scale magnetic
energy, Table \ref{tab:dynamo}), the magnetic field becomes
predominantly toroidal for stars rotating faster (up to 65\% of the
magnetic energy of HD~190771). The absence of any significant toroidal
component for slow rotators is unlikely to come from an intrinsic
limitation of ZDI in the low \vsin\ regime (where the sharp line
profiles may favour Zeeman signatures visible in the core of the line,
i.e. radial/meridional magnetic fields against the azimuthal field
component). The main reason, already invoked in sect. \ref{sect:energy},
is that the \vsin\ range of our sample is sufficiently limited to ensure
a consistent response of the imaging procedure over this domain. Another
confirmation of this idea comes from observations of \xib\ by Petit et
al. (2005),  where a predominant toroidal field is reported despite a
projected rotational velocity as low as 3 \kms\ (close to the \vsin\
estimate proposed by Valenti \& Fischer for HD~146233). 

The mainly poloidal geometries of HD~146233 and HD~76151 are reminiscent
of the large-scale solar magnetic geometry and might be typical of
low-activity dwarfs, at least for stellar masses close to the Sun's. For
higher spin rates, the development of a toroidal field is reminiscent of
previous ZDI studies of stars displaying a much higher magnetic activity
than the Sun (e.g. Donati et al. 2003, Petit et al 2005). Again, these
observational results are now supported by recent numerical simulations
of dynamo action in rapidly rotating solar-type stars (Brown et al.
2007b), where an increase by a factor 3 of the rotation rate compared to
the solar case results in a stable toroidal magnetic field at global
scales, throughout the bulk of the convective envelope and without the
help of any tachocline. The ordering of the
magnetic field onto a large-scale toroidal component when the star
rotation increases is usually
associated in MHD simulations with the inclusion of a
tachocline at the base of the convective zone (i.e. the addition of a strong
local shear between the differentially rotating convective zone
and the solid-rotating radiative core). Solar models
\textit{with} a tachocline lead to the building of strong
large-scale toroidal fields confined in this interface layer (Browning et al., 2006 \& 2007), but the
ability of these toroidal fields to travel the whole convective zone
while keeping their identity remains nevertheless a
matter of debate (e.g., Dorch \& Nordlund 2001, Jouve \& Brun 2007) and
the last simulations of rapidly rotating Suns \textit{without}
tachocline by Brown et al. precisely overcome this problem. From
our analysis, and despite the small number of objects in this first
available stellar sample, we infer that a rotation period smaller than
$\approx 12$~d is necessary for the toroidal component to reach an
energetic level similar to that of the large-scale poloidal field. This
is again in agreement with Brown et al. (2007b), where a dynamo model at
$\Omega = 3 \Omega_\odot$ produces a toroidal/poloidal field strength
ratio greater than unity.

\subsection{Energy splitting between low-order magnetic components}
\label{sect:complexity}

No clear rotational dependence is observed for the geometrical complexity of the reconstructed surface magnetic field. If we first consider the poloidal field only, a majority of the magnetic energy of this component is showing up in the dipole for \hdb\ and \hdc\ (with 79\% and 43\% of the total poloidal magnetic energy, respectively). For HD~146233, we reconstruct a predominantly quadrupolar field (56\% of the poloidal magnetic energy). For HD~73350, the poloidal magnetic energy is more evenly distributed within the dipole, quadrupole and octopole (with 24\%, 29\% and 33\% respectively). 

If we now consider the toroidal field alone (whenever its contribution is not negligible, i.e. for HD 73350 and HD 190771), we always measure a predominance of the mode $(l=1,~m=0)$ (87\% and 81\% of the total toroidal magnetic energy of HD 73350 and HD 190771, respectively). The large-scale toroidal field is therefore mainly organized as an axi-symmetric ring, displaying a striking similarity with the ``sea-snake" toroidal field structures produced by 3D MHD simulations in the presence of rapid rotation (Brown et al., 2007b).

In the Sun, the geometrical complexity of the large-scale field is varying during the solar cycle (e.g., Sanderson et al. 2003). An axisymmetric dipole is observed close to the solar minimum, but a high quadrupole/dipole ratio is observed at solar maximum. We can naturally expect time variability to occur for our stellar sample as well, implying that the energy splitting between low-order field components should be monitored over several years before discussing the possible informations it can disclose about the underlying dynamo.

\begin{table*}
\caption[]{Magnetic quantities derived from the set of magnetic maps. We list the mean unsigned magnetic field ($B_{\rm mean}$), the fraction of the large-scale magnetic energy reconstructed in the poloidal field component and the fraction of the {\em poloidal} magnetic energy in the dipolar ($\ell = 1$), quadrupolar ($\ell = 2$) and octopolar ($\ell = 3$) components. In the last column, we also list  $\log R'_{\rm HK}$ values derived from our sets of Stokes I spectra.}
\begin{tabular}{ccccccc}
\hline
Name & $B_{\rm mean}$ & pol. en. & dipole & quad. & oct. & $\log R'_{\rm HK}$  \\
  & (G) & (\% tot) & (\% pol) & (\% pol) & (\% pol) & \\
\hline
HD 146233 & $3.6\pm1$ & $99.3\pm0.2$ & $34\pm6$ & $56\pm6$ & $10\pm10$ & $-4.85\pm0.02$\\
HD 76151 & $5.6\pm2$ & $93\pm6$ & $79\pm13$ & $18\pm8$ & $3\pm3$ & $-4.69\pm0.02$\\
HD 73350 & $42\pm7$ & $52\pm3$ & $24\pm5$ & $29\pm8$ & $33\pm5$ & $-4.48\pm0.02$\\
HD 190771 & $51\pm6$ & $34\pm1$ & $43\pm8$ & $20\pm2$ & $23\pm4$ & $-4.42\pm0.02$\\
\hline
\end{tabular}
\label{tab:dynamo}
\end{table*}

\subsection{Differential rotation}
\label{sect:drot}

One of our target (HD 190771) possesses a measurable level of differential rotation. The absence of positive measurements on other targets reflects the combined effect of higher relative levels of noise and sparser phase sampling.  With a level of rotational shear about 2.3 times that of the Sun, HD 190771 is following nicely the mass-trend derived observationally by Barnes et al. (2005) for other rapidly-rotating stars (note that, probably because of this observational bias towards high rotation rates, the solar \drot\ parameters fall outside the mass dependence derived by Barnes et al.). The same authors also report an increase of \dom\ with \omeq\ (despite considerable scatter in available differential rotation measurements). Our own analysis is in global agreement with previous estimates derived for similar rotation rates by Doppler imaging (Barnes et al. 2005), broad-band photometry (Henry et al. 1995) and chromospheric monitoring (Donahue et al. 1996).  

Recent 3D numerical simulations investigate global flows in the presence of rapid rotation (Brown et al. 2007a). Their model simulate the convection up to 0.95 $R_\odot$ and ignore the very top layers of the convective zone. This radial limit could be problematic when comparing numerical simulations to observations, since the rotation can possibly undergo steep gradients just below the surface of rotating stars, as it does in the Sun (Schou et al. 1998). Whatever the conclusions of numerical simulations, we can therefore expect some difference with a surface estimate of differential rotation, as derived from photospheric magnetic maps. Despite this possible issue, Brown et al. (2007a) come to conclusions in good agreement with our observations, with a level of differential rotation \dom\ of 1 M$_\odot$  stars increasing with the spin rate by a factor of $\approx$3 when increasing the rotation from 1 to 3 $\Omega_\odot$. 

Our isolated measurement of \drot\ clearly deserves to be repeated on other stars, to confirm the shear level we observe at 3$\Omega_\odot$ for HD 190771 and derive similar measurements for different rotation rates. From our analysis, we can already propose that an increase of \dom\ from slow to fast rotation may contribute to make the $\Omega$ effect of the stellar dynamo more efficient and may, as a consequence, help to build a global-scale toroidal field throughout the convective zone.

\subsection{Long-term magnetic variability}
\label{sect:var}

The various magnetic properties we derive here may be subject to long-term variability. The monitoring of chromospheric emission reported by Baliunas et al. (1995) and Hall et al. (2007a, 2007b), actually shows that two of our targets (HD 76151 and HD 146233) follow activity cycles. 

With a period of 7 years (Hall et al. 2007b), the cycle length of HD 146233 is shorter than the solar cycle. The chromospheric activity of this solar twin was reported to sharply grow between 2004 and 2006 (after a regular decrease between 2000 and 2004), so that our spectropolarimetric observations of the summer of 2007 are likely to be representative of a high activity state. The predominantly quadrupolar magnetic geometry we reconstruct at that time is reminiscent of the magnetic topology of the Sun at solar maximum. However, despite the fact that HD 146233 is known as the best solar twin among bright stars, its short cycle length indicates that any comparison with the solar magnetic topology should be considered with caution. 

The long-term chromospheric variability of HD 76151 (Baliunas et al., 1995) suggests the existence of a 2.5 yr cycle (of weak amplitude compared to chromospheric solar variability), superimposed on a possibly much longer cycle, of higher amplitude and unknown period. Its last activity peak was recorded in 2006 (Hall et al. 2007a), so that our data set was collected during the following decrease of activity. 

No such long-term chromospheric follow-up is available for HD 190771 and HD 73350. The comparison with other stars of similar activity levels and spectral types (Baliunas et al. 1995) reveal however that the cyclicity undergone by their CaII emission is generally less pronounced than in the Sun, with a range of possible cycle lengths (depending on the star) and the co-existence of different time-scales of variability.

\section{Conclusions}

From a set of stellar spectropolarimetric observations, we report the detection of surface magnetic fields in a sample of four solar-type stars. Assuming that the observed variability of the polarimetric signal is controlled by stellar rotation, we establish the rotation periods of our targets, with values ranging from 8.8~d (for \hdc) to 22.7 d (for \sco). One star of our sample (HD 190771) displays a measurable level of surface differential rotation, with a rotational shear about 2.3 times stronger than solar. Apart from rotation, fundamental parameters of the selected objects are very close to the Sun's, making this sample a practical basis to investigate the specific impact of rotation on magnetic properties of Sun-like stars.

We reconstruct the large-scale magnetic geometry of the targets as a low-order ($\ell<10$) spherical harmonics expansion of the surface magnetic field. From the set of magnetic maps, we draw two main conclusions. (a) The magnetic energy of the large-scale field increases with rotation rate. The increase of chromospheric emission with the mean magnetic field is flatter than observed in the Sun. Even if we cannot completely rule out the possible influence of 
increased spottedness in more active stars, this observation may suggest that a larger fraction of the surface magnetic energy is stored in large scales as rotation increases (since the chromospheric flux is also sensitive to magnetic elements smaller than those contributing to the polarimetric signal). (b) Whereas the magnetic field is mostly poloidal for low rotation rates, more rapid rotators host a large-scale toroidal component in their surface field. From our observations, we infer that a rotation period lower than $\approx$12 days is necessary for the toroidal magnetic energy to dominate over the poloidal component. 

In future work, we will investigate how this rotation threshold may vary for various stellar masses, and therefore study how the depth of the convection zone may influence the existence of global-scale toroidal magnetic fields at the stellar surface. This enlargement of the available stellar sample is already under way. The magnetic properties we derive here may also be subject to some level of time variability. This idea is supported by the long-term monitoring of chromospheric emission of Baliunas et al. (1995) and Hall et al. (2007a), in which two of our targets  (HD 76151 and HD 146233) are reported to follow activity cycles. Variabilities of the large-scale magnetic geometries related to such fluctuating activity remain to be unveiled. A long-term observing program is now started to gather the spectropolarimetric data sets necessary to tackle this question, with the objective to offer a completely new set of observables with which to constrain stellar dynamo theories.

\begin{figure}
\centering
\includegraphics[width=9cm]{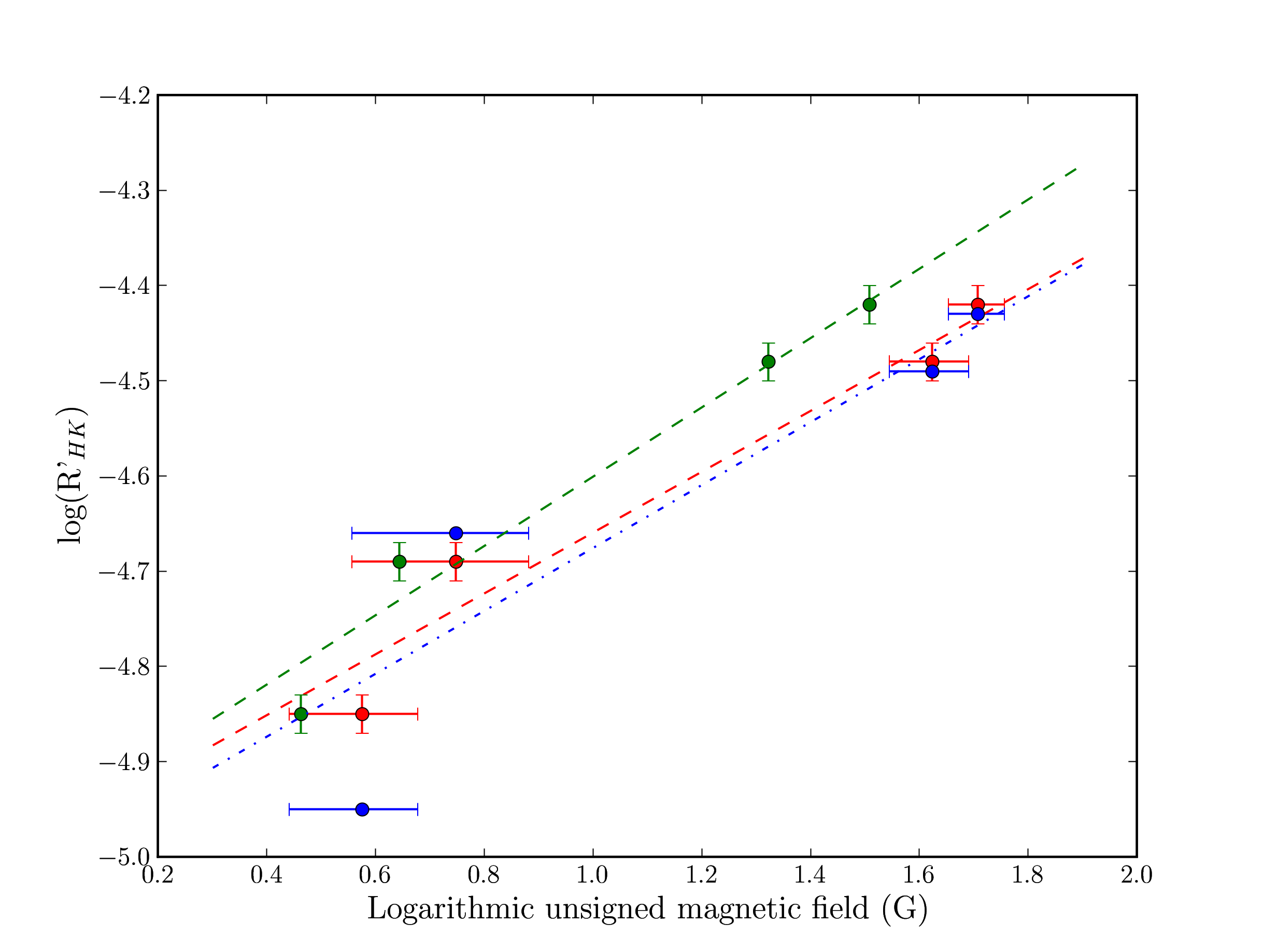}
\caption{Chromospheric flux ($\log R'_{HK}$) as a function of the large-scale logarithmic magnetic field. Chromospheric data plotted as blue dots are taken from Wright et al. (2004) and Hall et al. (2007a). Red dots represent measurements of $\log R'_{HK}$ derived from our own sets of Stokes I NARVAL spectra. The green dots are also obtained from NARVAL observations of CaII emission, but they are plotted against the {\em radial} field component extracted from the magnetic maps. The dashed lines are the result of power-law adjustments discussed in section 4.1.}
\label{fig:rhk}
\end{figure}

\begin{figure}
\centering
\includegraphics[width=9cm]{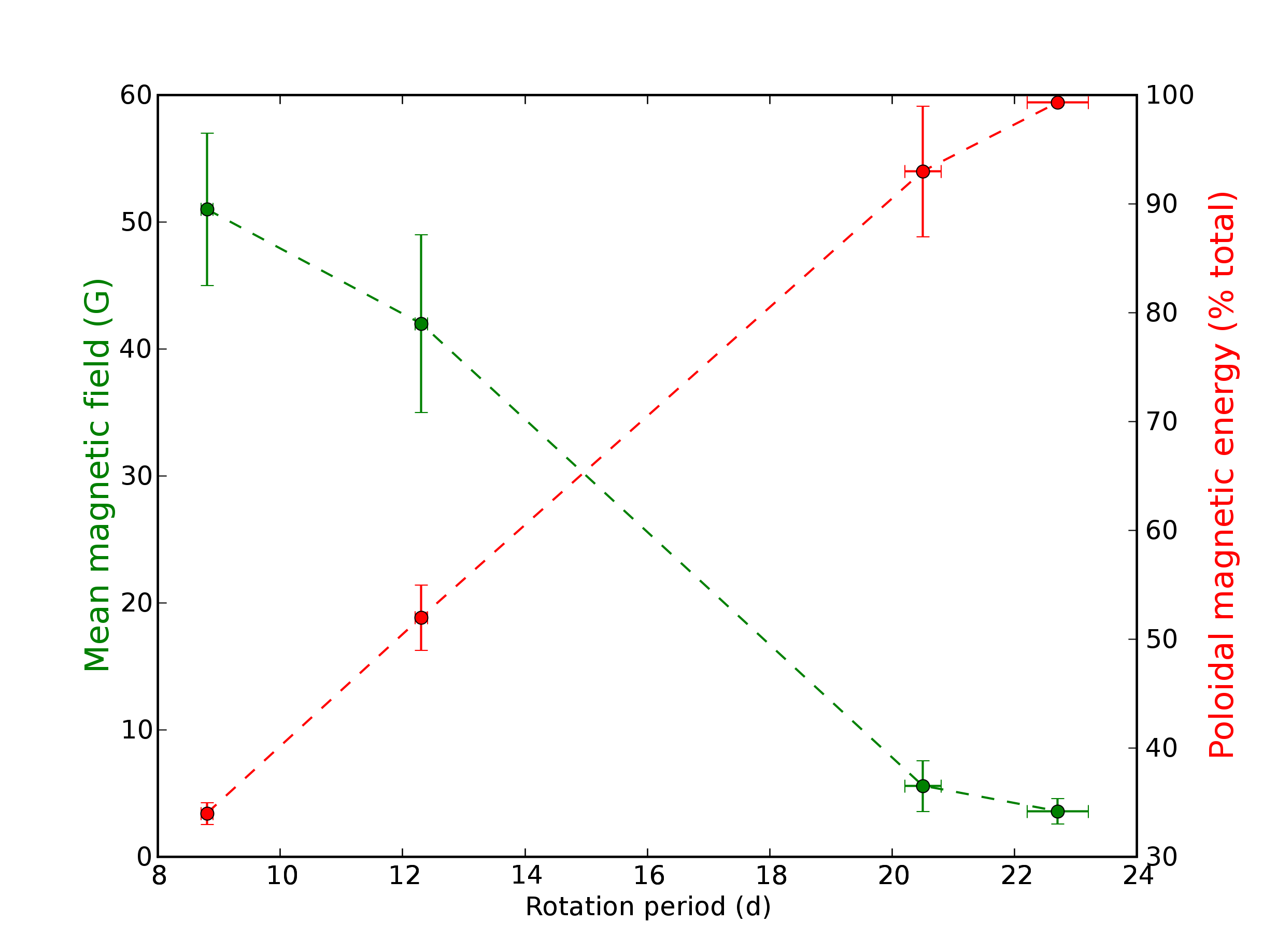}
\caption{Rotational dependence of the mean (unsigned) magnetic field (green line) and of the fraction of magnetic energy stored in the poloidal field component (red curve).}
\label{fig:energy}
\end{figure}

\section*{ACKNOWLEDGMENTS}

We thank the staff of TBL for their help during the observing runs. We acknowledge use of the SIMBAD and VizieR data bases operated at CDS, Strasbourg, France. We are grateful to an anonymous referee for useful comments that helped to clarify and improve this article.

\end{document}